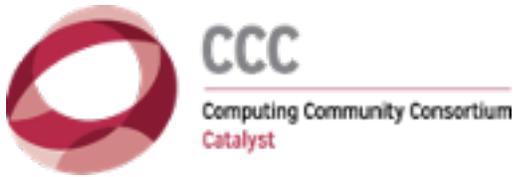

# Taking Stock of the Present and Future of Smart Technologies for Older Adults and Caregivers

*A Computing Community Consortium (CCC) Quadrennial Paper*
Christina N. Harrington (DePaul University), Ben Jelen (Indiana University), Amanda Lazar (University of Maryland), Aqueasha Martin-Hammond (Indiana University Purdue University - Indianapolis), Alisha Pradhan (University of Maryland), Blaine Reeder (University of Missouri) Katie Siek (Indiana University)

Technology has the opportunity to assist older adults as they age in place, coordinate caregiving resources, and meet unmet needs through access to resources. Currently, older adults use consumer technologies to support everyday life, however these technologies are not always accessible or as useful as they can be. Indeed, industry has attempted to create smart home technologies (e.g., Microsoft HomeOS, Intel CareNet) with older adults as a target user group, however these solutions are often more focused on the technical aspects and are short lived. In this paper, we advocate for older adults being *involved* in the design process - from initial ideation to product development to deployment. We encourage federally funded researchers and industry to create compensated, diverse older adult advisory boards to address stereotypes about aging while ensuring their needs are considered.

We envision artificial intelligence (AI) systems that augment resources instead of replacing them - especially in under-resourced communities. Older adults rely on their caregiver networks and community organizations for social, emotional, and physical support; thus, AI should be used to coordinate resources better and lower the burden of connecting with these resources. Although sociotechnical smart systems can help identify needs of older adults, the lack of affordable research infrastructure and translation of findings into consumer technology perpetuates inequities in designing for diverse older adults. In addition, there is a disconnect between the creation of smart sensing systems and creating understandable, actionable data for older adults and caregivers to utilize. We ultimately advocate for a well-coordinated research effort across the United States that connects older adults, caregivers, community organizations, and researchers together to catalyze innovative and practical research for *all* stakeholders.

### Involving Older Adults in the Design of Systems
*Older adults use mobile phones, voice assistants, and apps*, however technology companies may not include older individuals in their design process, or as employees. Thus, these technologies are designed for younger people's eyes, fingers, experiences, etc. For example, advanced digital technologies such as fingerprint recognition are tested with young adults and do not work with older adults because older adults tend to have more wrinkles on their skin – making it difficult to recognize their fingerprint. There needs to be a continuous push to include older adults in all technical design processes – from needs assessments to usability testing for *all technology*, not just the technology designed specifically for older adults. *Technology companies should have advisory boards that include older adults in setting priorities for technology research and development.* Diverse representation on advisory boards helps with misperceptions of what it is like to be an older adult, where younger people may assume that aging is all about decline (and consequently design technology to alleviate cognitive and physical decline) and not realize the ways that older people want to be involved in being active members of their communities. Individuals with different backgrounds, identities, abilities, needs, and characteristics will have different priorities and therefore it is key to work with diverse older adults in setting priorities for technologies.

Research has found that *older adults use and react to technologies in ways that designers do not always expect*. For example, some people may not wear an emergency alert pendant because it is stigmatizing or reminds them of mortality - this is an example of where form factor is important in design. In other cases, the setup that a technology requires may not be feasible even if the technology itself is easy to use. Industry and academic researchers must investigate technology use (and non-use) to identify the intended effects of technology.

When older adults are engaged in usability studies, design teams should recruit a broad sample. Convenience sampling may not be the most effective because often those who are readily interested and able to participate in research studies have access or knowledge about the institution or organization conducting research. *There is a need to make our research practices themselves more inclusive to reach the margins of the aging population.* In many cases, technology is being designed and developed that only considers samples of older adults who have access and familiarity with technology and access to the internet as a platform for technology solutions. As a result of this, lower-income, and ethnic minorities, who may experience lower digital literacy, are left out of consideration of newer, more innovative approaches to supporting health, functioning, socialization and connection. As an example, lower-income, ethnic minority older adults are found to feel less capable of engaging voice-controlled smart devices, such as conversational assistants, in self-managing healthcare in the home. One promising area of research from the University of Michigan is where older adults can access digital resources through their telephones[1]. Another challenge in our landscape is shifting from considering older adults who are not White, digitally proficient, or affluent as "special populations" instead of centralizing them as the focus of our research and development efforts. *Researchers and industry should report demographics on who helped to inform, study, and evaluate the systems they market to the general public.*

Outside of the design and evaluation process, *older adults must be included in training data sets that inform intelligent system models.* Smart home appliances with embedded AI are often advertised as ideal technology to help older adults age in place. Older adults can benefit from the social and companionship features some of these systems provide, however recent studies have shown major usability issues in smart voice assistants that do not wait for older adults to complete their command before the device starts the interaction - thus, the older adult becomes frustrated and the system cannot successfully complete the command. We must better understand the social roles older adults want these systems to play in their lives and ensure the algorithms that dictate interactions accommodate older adult interactions.

---

[1] https://sph.umich.edu/news/2020posts/virtual-senior-center-helps-older-adults-detroit-connect.html

**Balancing Assistance and Social Needs**

Advancements in artificial intelligence have brought new opportunities for supporting older adults as they age in place. Industry and academic researchers are exploring everything from social robots to voice assistants that provide older adults with companionship. They are developing new tools to streamline healthcare tasks at home and in clinical settings. However, as AI advances, we must also be careful to understand *how AI-enabled technologies meaningfully fit into older adults' daily lives, including how they impact social interactions, relationships, and connections with others.* In the context of care, connecting with people can be just as important as meeting physical needs. For some older adults, engaging with formal and informal caregivers can address basic needs for social interaction. Even technology-enabled interactions, such as Telecare, can serve as a form of companionship. *Thus, for some, replacing caregiving tasks with technology may mean that needs for social interaction go unmet.* Although AI might augment older adults' lives, we need to understand how to *balance AI's use for addressing physical needs with other social dimensions of care*.

AI has the potential to streamline tasks and make them more efficient, however, many caregivers and organizations spend years crafting their expertise and building rapport and trust with the people and communities they serve. The specialized knowledge generated may not easily be transferred to a digital environment. This lack of transferability can bring to question the appropriateness of AI-enabled technologies for different care-related tasks. Working with healthcare professionals, other service providers, and organizations in the community (e.g., Indiana's Areas on Aging organizations) can help. Older adults may also have informal support networks within their communities (e.g., friends, neighbors). *Leveraging opportunities for human-AI collaboration that integrate human expertise in AI-enabled interactions may help address emerging concerns related to the trust, privacy, and safety of these technologies.*

In the past year, the pandemic has heightened awareness of the digital inequalities experienced by those who do not have access to the essential technical resources needed to study, work, and receive care. This is especially apparent in the challenges older adults face in accessing online information such as finding vaccine locations and scheduling appointments. Therefore, even as AI advances, there are some that are not benefiting from those advancements due to access barriers. *For those older adults without access to the Internet, computers, or other digital resources, it will be vital to prevent further digital inequalities by including people as a part of AI-enabled interactions and carefully measuring the impact of possible less human interaction with older adults.*

**The Role of Smart Environments**

A primary challenge to supporting equity in technical systems that enable aging in place is the lack of affordable infrastructure. Major contributors to lack of infrastructure are the traditional model of research funding and market-driven solutions that place affordable access to technology beyond the reach of older adults. The nature of research funding is project-based and term-limited. *While this funding model excels at driving innovations, it provides no mechanism for persistent access to promising technology-based services developed through research*. As a result, project momentum is interrupted while technology researchers search for new funding options and fruitful collaborations may terminate because funding is not sustained. Further, technology innovations that show potential to support aging in place can end up beyond the reach of older adults as a result of the commercialization process. Technology startups that attempt to commercialize these innovations face pressures to remain viable in the marketplace.

*Lack of infrastructure impedes translational research to understand what technology innovations will work if they are implemented for longer periods outside of lab settings or term-limited research projects*. In addition, the infrastructure gap also impedes efforts to enhance evidence-based programs. For example, the CAPABLE program developed at Johns Hopkins University, is a multi-disciplinary home visit program delivered by occupational therapists and nurses that has shown significant reductions in disability for older adults living at home. Programs like CAPABLE could be enhanced using sensor technology and telehealth to extend reach to older adults who live in rural areas and send timely alerts of functional decline to avoid falls and hospitalizations. Currently, we cannot conduct wide-scale deployments of evidence-based solutions because we do not have the affordable infrastructure to develop and evaluate these technology approaches, nor do some of these areas have internet access.

*Non-profit organizations with wide community input have shown promise in helping to translate research to consumer-grade technologies to help older adults*. One example of a non-profit effort is Coach Me Health (https://www.coachmehealth.org/), a non-profit technology company whose mission is "to improve the lives of low-income Americans on Medicaid with personalized, culturally-sensitive, and tech-enabled care." Another non-profit effort to provide persistent support for academic research projects until they are ready to scale to production is Hekademeia Research Solutions (https://www.hekademeia.org/). These non-profit efforts however are still in an emergent phase where sustainability models are being developed and tested. Indeed, another non-profit, All Mental Health (https://www.allmentalhealth.org/) recently merged with Coach Me Health after facing financial difficulty. All Mental Health was formed to maintain access to technology that was demonstrated as effective in funded research but was the intellectual property of a failed startup (Lantern). Utilizing non-profits to help translate research to broader access stands in contrast to the current approach of subsidizing technology access by allocating funds to public technology companies. An additional challenge to making this translation a reality is that current SBIR/STTR awards cannot be directly received by non-profits.

### Machine Learning and AI to Support Caregiving
Computing specialists have utilized sensors, computer networks, and artificial systems to assist older adults safely age in place. Essentially, these are tracking systems that can do everything from tell where an older adult is and alert someone if they wander out of a specific geographical location to send alerts to help older adults take their medication on schedule. When a system has enough data, it can also detect declines in older adults to assist stakeholders with assessing support and infrastructure needs.

Currently, more research is needed in the sensors and infrastructure used to sense these declines to improve the data quality. We do not want to use inaccurate data that could deeply impact the care and financial resources of an older adult. These technological advances can help older adults live safer, independent lives; however, we know less about how this type of information should be delivered. *We need interdisciplinary approaches to investigate how findings on one's decline can be verified and provided to older adults and their caregivers in an understandable and actionable format*.

Machine learning and Artificial Intelligence has traditionally been used to detect anomalies in older adult lives to help caregivers feel better about their older adult loved one's safety. Did an older adult loved one move around today? Did anyone visit? Did they fall? Rich research continues in these areas as computing researchers learn that they must better understand one's context and social networks to make accurate determinations (e.g., Your adult loved one got up repeatedly for two hours at uncertain intervals because it was Halloween and they were giving treats to children; no one visited despite the door opening). *We must continue to investigate what can be done with all of this information and inferences with all stakeholders.*

One area not well investigated is utilizing artificial intelligence to coordinate caregiving, social support, community resources (that are often siloed), and healthcare systems. Typically, this coordination task is done by one person in an older adult's care network. In addition, although people offer to help, understanding what help people can give and when is challenging. As we move towards more networked and digital forms of scheduling and resource management, future AI systems could *negotiate these care systems to assist getting older adults the resources they need with limited caregiver overhead*.

### Recommendations for Moving Forward

**1. Creating Persistent Knowledge and Community Ties through Research Centers.** We propose building a network of federally funded research centers across the country that can build collaborations with local older adult communities and non-profits. The centers would develop resources that can serve as a standard practice for collaborating researchers to utilize and integrate into their own research practices. The center can facilitate collaboration between researchers around the country so that research can be assessed from a diverse group of older adults instead of a local, convenient sample.

2. **Integrating Feedback from Older Adults.** Industry and academic researchers should engage older adults throughout the design process for technology designed for the general population. Requiring advisory boards of diverse older adults for federally funded grants and corporations is one step to ensuring older adults have a voice at the table. Researchers should demonstrate how advisory board feedback was integrated into the design and final deliverables. The national network of research centers can assist researchers and industry identify advisory board participants, thus creating a national forum for older adults to participate in.

3. **Building Accountability into Federally Funded Research.** There must be better efforts to ensure that the older adults who participate in the research that shapes technology mirrors the populations potentially affected by the technology. Ensuring participants' diversity in federally funded research through longitudinal reporting and accountability could motivate inclusive recruitment practices and critical reflection of how the technology impacts groups of older adults with different lived experiences.

4. **Supporting Research Equity and Diversity.** In some cases, more federal funds should be allocated to help build and maintain relationships with community groups, especially those in underserved communities, who may not have the resources. Diversification supplements could be used at the time of grant request in the solicitation to help researchers work on innovative, transformative research, while also ensuring the diverse communities they work with have the resources and infrastructure necessary (e.g., internet) to benefit from the products.

5. **Innovating Beyond Technical Solutions.** Technology researchers have made great advancements. However, we must shift our goal from technical innovation alone if we want our research to work within the complex realities of people's lives. Advancing the science of implementation, addressing usability issues, and supporting technology access and infrastructure should all be seen as key goals for funders, technology innovators, and scientific progress.

### Disclosures
Blaine Reeder is Co-Founder and Chair of the Board of Directors of Hekademeia Research Solutions, a 501(c)(3) nonprofit organization. Dr. Reeder receives no direct financial benefit from this relationship. He is a user of Hekademeia's services.

### Acknowledgements
We would like to thank Pallabi Bhowmick for her brainstorming on this article.

*This white paper is part of a series of papers compiled every four years by the CCC Council and members of the computing research community to inform policymakers, community members and the public on important research opportunities in areas of national priority. The topics chosen represent areas of pressing national need spanning various subdisciplines of the computing research field. The white papers attempt to portray a comprehensive picture of the computing research field detailing potential research directions, challenges and recommendations.*

*This material is based upon work supported by the National Science Foundation under Grant No. 1734706. Any opinions, findings, and conclusions or recommendations expressed in this material are those of the authors and do not necessarily reflect the views of the National Science Foundation.*

*For citation use: Harrington C., Jelen B., Lazar A., Martin-Hammond A., Pradhan A., Reeder B., and Siek K. (2021) Taking Stock of the Present and Future of Smart Technologies for Older Adults and Caregivers. https://cra.org/ccc/resources/ccc-led-whitepapers/#2020-quadrennial-papers*